\def\beq{\begin{equation}}
\def\eeq{\end{equation}}

\def\virg#1{``#1''}

\documentclass{ws-ijmpd}
\usepackage{latexsym}

\begin{document}

\markboth{L. Iorio \& M. L. Ruggiero}
{Phenomenological Constraints on Ho\v{r}ava-Lifshitz Solution}

%
%

\title{PHENOMENOLOGICAL CONSTRAINTS ON THE\\ KEHAGIAS-SFETSOS SOLUTION IN THE\\ HO\v{R}AVA-LIFSHITZ GRAVITY FROM\\ SOLAR SYSTEM ORBITAL MOTIONS }

\author{L. IORIO}

\address{Ministero dell'Istruzione, dell'Universit\`{a} e della Ricerca (M.I.U.R.). Permanent address: Viale Unit\`{a} di Italia 68\\
Bari, (BA) 70125,
Italy\\
lorenzo.iorio@libero.it}

\author{M. L. RUGGIERO}

\address{
Dipartimento di Fisica, Politecnico di Torino, Corso Duca degli Abruzzi 24\\  Torino, (TO) 10129, Italy\\
matteo.ruggiero@polito.it}

\maketitle

\begin{history}
\received{3 July 2010}
\end{history}

\begin{abstract}
We focus on  Ho\v{r}ava-Lifshitz (HL) theory of gravity, and, in particular, on the   Kehagias and Sfetsos's  solution that is the analog of Schwarzschild black hole of General Relativity.
In the weak-field and slow-motion approximation we analytically work out the secular precession of the longitude of the pericentre $\varpi$ of a test particle induced by this solution. Its analytical form is different from that of the general relativistic Einstein's pericentre precession. Then, we compare it to the latest determinations of the corrections $\Delta\dot\varpi$ to the standard Newtonian/Einsteinian planetary perihelion precessions recently estimated by E.V. Pitjeva with the EPM2008 ephemerides. It turns out that the  planets of the solar system, taken singularly one at a time, allow to put lower bounds on the adimensional HL parameter $\psi_0$ of the order of $10^{-12}$ (Mercury)$-10^{-24}$ (Pluto). They are not able to account for the Pioneer anomalous acceleration for $r>20$ AU.
\end{abstract}


\section{Introduction} \label{sec:intro}
Ho\v{r}ava\cite{horava1,horava2} has recently proposed a four-dimensional theory of gravity that can be thought of as a candidate for a UV completion of General Relativity (GR), since it is power-counting renormalizable. This theory admits the Lifshitz  scale invariance: $\mathbf{x} \rightarrow b \mathbf{x}, \quad t \rightarrow b^{q} t$, and, after this, it is referred to as Ho\v{r}ava-Lifshitz (HL) theory. It is  $q=3$ in the UV,  and hence HL theory exhibits anisotropy between space and time, while relativistic scaling with  $q=1$ is recovered in the IR regime. Actually, in the original formulation of HL gravity there are problems in recovering GR at large distances, unless the condition of ``detailed balance'' (used by Ho\v{r}ava to restrict the number of possible parameters) is relaxed\cite{pope,KS}. Thus, phenomenologically viable modifications\cite{visser1,visser2} of the theory have been considered, even though some problems still seem to be present\cite{padilla,blas}.

Since its formulation, HL gravity has been deeply investigated; several of its applications (e.g. cosmology, study of exact solutions, lensing, black holes and thermodynamics of black holes) have been analyzed (see Refs.~\refcite{chen,kon} and references therein). In particular, a static, spherically symmetric solution has been found\cite{KS} by Kehagias and Sfetsos  (KS). Such a solution is the analog of Schwarzschild one of GR and, moreover, it asymptotically reproduces the usual behavior of Schwarzschild spacetime. The KS solution is obtained without requiring the projectability condition, assumed in the original HL theory. Spherically symmetric solutions with the projectability condition are, however, available\cite{tang,Wang,Greenwald}. It is important to point out that, up to now, in HL gravity the gravitational field is purely geometrical: in other words, the way matter has to be embedded still needs to be studied. Furthermore, since the way matter couples to gravity is not clear, the fact that particles move along geodesic is not granted, and deviations from geodesic motion are expected\cite{capasso}. That being said, it is possible to consider the  KS solution as a toy model which can be studied to obtain a better theoretical insight into the fundamental aspects of HL gravity  because of its relative simplicity.  Actually, this has already been done, focusing on implications of this solution in various physical scenarios, ranging from accretion disks\cite{Harko} to black holes thermodynamics (see Ref.~\refcite{Janke} and references therein).

Since the Newtonian and lowest order post-Newtonian limits of the KS solution coincides with those of GR, it is manifest that this solution of HL gravity is in agreement with the classical tests of GR, as it has recently been shown in Ref.~\refcite{Tib}. In this paper we focus on the first corrections to the GR behavior, in order to evaluate the impact of these corrections on  solar system orbital motions. Furthermore, we use data from solar system observations to constrain the KS solution. In particular,  we want to investigate the weak-field and slow-motion approximation of the KS solution  by working out the induced pericenter  precession of a test particle (Section \ref{calcolo}) and comparing it to recent observations-based determinations of the non-standard rates of the perihelia of some planets of the solar system (Section \ref{paragone}). Section \ref{concludi} is devoted to the discussion and the conclusions.
\section{Calculating the perihelion precession}\label{calcolo}
In order to study orbital motion in the gravitational field of the Sun, we start from a static and spherically symmetric  metric
\beq ds^{2}= e^{\nu(r)}c^2 dt^2 - e^{\lambda(r)}dr^2 -r^2d\Phi^2,\eeq
with $d\Phi^{2}=d\vartheta^{2}+\sin^{2} \vartheta d\varphi^{2}$, in Schwarzschild-like coordinates.
In particular, for the asymptotically flat   KS solution we have\cite{KS}
\beq e^{\nu(r)}=e^{-\lambda(r)}=1+\psi r^{2}-\sqrt{\psi^{2}r^{4}+4 \psi \mathcal{M} r}, \label{eq:metricas} \eeq with
$\mathcal{M}\equiv\frac{GM}{c^2}$, so that
\beq
ds^{2}=\left(1+\psi r^{2}-\sqrt{\psi^{2}r^{4}+4 \psi \mathcal{M} r} \right)c^2 dt^{2}-\frac{dr^{2}}{\left(1+\psi r^{2}-\sqrt{\psi^{2}r^{4}+4 \psi \mathcal{M} r} \right)}-r^{2}d\Phi^{2}. \label{eq:metric1}
\eeq
Let us re-write
$e^{\nu(r)}$ as
\beq e^{\nu(r)}=1+\psi r^2\left[1-\sqrt{1+\frac{4\mathcal{M}}{\psi r^3}}\right].\eeq
Let us focus on a typical Sun-planet system so that $r\approx 1$ AU; it is clear that some limiting condition on $\psi$ must hold  so that
\beq e^{\nu(r)}\rightarrow 1-\frac{2\mathcal{M}}{r},\eeq as in the  Schwarzschild case. After posing $\psi \doteq \frac{\psi_0}{\mathcal{M}^2}$, it turns out that it must be
\beq \psi_0\gg 4\left(\frac{\mathcal{M}}{r}\right)^3\approx 4\times 10^{-24}\label{condiz}\eeq for $M=M_{\odot}$ and $r\approx 1$ AU. Indeed, in this case, one has
\beq e^{\nu(r)}\approx 1-\frac{2\mathcal{M}}{r} + \frac{2}{\psi_0}\left(\frac{\mathcal{M}}{r}\right)^4,\eeq so that
\beq
ds^{2}=\left(1-\frac{2\mathcal{M}}{r}+\frac{2\mathcal{M}^{4}}{\psi_0 r^{4}}\right)c^2 dt^{2}-\frac{1}{\left(1-\frac{2\mathcal{M}}{r}+\frac{2\mathcal{M}^{4}}{\psi_0 r^{4}}\right)}dr^{2}-r^{2}d\Phi^{2}. \label{eq:metric2}
\eeq
Even though here we are concerned on the KS solution of HL gravity, our approach apply as well to an arbitrary perturbation of the Schwarzschild solution in the form
\beq
e^{\nu(r)}\approx 1-\frac{2\mathcal{M}}{r} + \frac{2\alpha}{r^{4}}, \label{eq:alpha1}
\eeq
where $\alpha$ is a suitable perturbation parameter: in fact, on setting $\psi_{0} = \mathcal{M}^{4}/ \alpha$, it is possible to get constraints on $\alpha$ starting from the ones obtained on $\psi_{0}$, knowing the mass of the source $M=c^{2}\mathcal{M}/G$.

By inspection of the metric (\ref{eq:metric2}), it is clear that the first $\psi_0-$type corrections to the GR behaviour for large distances are of order $r^{-4}$; We aim at investigating the effects of these corrections that depend on terms in the form $\frac{\mathcal{M}^{4}}{\psi_0 r^{4}}$ and we neglect the higher order effects of Schwarzschild metric (i.e. the terms that are proportional to $\left(\frac{\mathcal{M}}{r}\right)^{k},  k\geq 2$).

To this end, we introduce the isotropic radial coordinate $\bar r$
\beq
\bar r = r \left (1-\frac{\mathcal{M}}{r}+\frac{1}{4}\frac{\mathcal{M}^{4}}{\psi_0 r^{4}}\right) \label{eq:defrbar}
\eeq
and substitute in (\ref{eq:metric2}). After approximating, the metric becomes
\beq
ds^{2}=\left(1-\frac{2\mathcal{M}}{\bar r}+\frac{2\mathcal{M}^{4}}{\psi_0 \bar r^{4}} \right)c^2 dt^{2}-\left(1+\frac{2\mathcal{M}}{\bar r}-\frac{1}{2}\frac{\mathcal{M}^{4}}{\psi_0 \bar r^{4}} \right) \left(d\bar r^{2}+\bar r^{2}d\Phi^{2}\right) \label{eq:metric22}
\eeq
or, introducing Cartesian coordinates such that $\bar r^{2}=x^{2}+y^{2}+z^{2}$,
\beq
ds^{2}=\left(1-\frac{2\mathcal{M}}{\bar r}+\frac{2\mathcal{M}^{4}}{\psi_0 \bar r^{4}} \right)c^2 dt^{2}-\left(1+\frac{2\mathcal{M}}{\bar r}-\frac{1}{2}\frac{\mathcal{M}^{4}}{\psi_0 \bar r^{4}} \right) \left(dx^{2}+dy^{2}+dz^{2}\right); \label{eq:metric3}
\eeq
in the following we will re-name $\bar r$ as $r$.\footnote{It useful to remember that the isotropic form of Schwarzschild metric is, up to $\left(\frac{\mathcal{M}}{r}\right)^{2}$,
\[
ds^{2}=\left(1-\frac{2\mathcal{M}}{r}+\frac{2\mathcal{M}^{2}}{ r^{2}} \right)c^2 dt^{2}-\left(1+\frac{2\mathcal{M}}{ r}+\frac{3}{2}\frac{\mathcal{M}^{2}}{ r^{2}} \right) \left(d r^{2}+ r^{2}d\Phi^{2}\right). \label{eq:metricSchiso}
\]}

From eq.(\ref{eq:metric3}) and from the post-Newtonian equations of motion of  a test particle\cite{Brum}
\beq \ddot x^i = -\frac{1}{2}c^2 h_{00,i} - \frac{1}{2}c^2 h_{ik}h_{00,k}+h_{00,k}\dot x^k\dot x^i + \left(h_{ik,m}-\frac{1}{2}h_{km,i}\right)\dot x^k\dot x^m,\ i=1,2,3,\eeq
valid for time-independent space-time metrics without off-diagonal terms,
it is possible to obtain the following expression for the $\psi_0-$dependent acceleration
\beq \vec{A}_{\psi_0}=\frac{G^4 M^4}{\psi_0 c^6 r^5}\left[\left(4+\frac{v^2}{c^2}\right)\hat{r} -10\left(\frac{v_r}{c^2}\right)\vec{v}\right],\label{acce}\eeq
where $\hat{r}\doteq\frac{\vec{r}}{r}$ is the planet's unit vector in the radial direction, $\vec{v}$ is the planet's velocity, and $v_r$ is its component in the radial direction, i.e. $v_r\doteq\vec{v}\cdot \hat{r}$. Since typical solar systems's planetary speeds are of the order of $v\approx na\approx 10^4$ m s$^{-1}$, where $a$ is the semi-major axis and $n\doteq\sqrt{GM/a^3}$ is the Keplerian mean motion related to the orbital period  by $P_{\rm b} \doteq 2\pi/n$, we can neglect the terms $\frac{v^2}{c^2}$ and, especially, $\left(\frac{v_r}{c^2}\right)v$, which is even smaller since $v_r$ is proportional to the planet's eccentricity $e$\cite{Roy}. Thus, $\vec{A}_{\psi_0}$ reduces to the entirely radial term
\beq \vec{A}_{\psi_0}\approx \frac{4 G^4 M^4}{\psi_0 c^6 r^5}\hat{r}\eeq whose magnitude, in the case $M=M_{\odot}$ and $r\approx 1$ AU, is quite smaller than the Newtonian monopole for $\psi_0>4\left(\frac{\mathcal{M}}{r}\right)^3$; as we will see later, such a condition is fully satisfied in the solar system. As a consequence, it is possible to apply the standard Gauss perturbative approach\cite{Ber}
to derive the secular, i.e. averaged over one orbital period, precession of the longitude of the perihelion\cite{Roy} $\varpi\doteq\Omega+\omega$, where $\Omega$ is the longitude of the ascending node and $\omega$ is the argument of the perihelion.
The Gauss equation for the node's variation is\cite{Ber}
\beq \dot\Omega = \frac{1}{na\sqrt{1-e^2}\sin I}A_n\left(\frac{r}{a}\right)\sin (\omega + f),\label{nodo}\eeq where $A_n$ is the component of the perturbing acceleration $\vec{A}$, whatever its physical origin may be, normal to the orbital plane, $I$ is the inclination\footnote{$\Omega,\omega,I$ can be thought as the three Euler angles determining the spatial orientation of a rigid body, i.e. the un-perturbed Keplerian ellipse which changes neither its size, fixed by $a$, nor its shape, fixed by $e$, with respect to a locally quasi-inertial frame.} to the ecliptic, and $f$ is the true anomaly reckoning the planet's position from the the perihelion\cite{Roy}. The Gauss equation for the variation of $\omega$ is\cite{Ber}
\beq\dot\omega = \frac{\sqrt{1-e^2}}{nae}\left[-A_r \cos f +A_{\tau}\left(1+\frac{r}{p}\right)\sin f\right]-\cos I\dot\Omega,\label{perielio}\eeq where $p\doteq a(1-e^2)$ is the semi-latus rectum, and $A_r$ and $A_{\tau}$ are the components of $\vec{A}$ along the radial and transverse directions, respectively, both in-plane.
The un-perturbed Keplerian ellipse, on which the right-hand-sides of eq. (\ref{nodo}) and eq. (\ref{perielio}) have to be evaluated, is\cite{Roy}
\beq r = \frac{a(1-e^2)}{1+e\cos f}, \eeq while the average over one orbital period has to be performed by means of\cite{Roy}
\beq dt = \frac{(1-e^2)^{3/2}}{n(1+e\cos f)^2} df.\eeq
Since, in this case, there is no normal component of $\vec{A}_{\psi_0}$, the node is left unaffected. The calculations with eq.(\ref{perielio}) are made simpler by the absence of the transverse component of $\vec{A}_{\psi_0}$ as well. As a consequence, we obtain that the advance of the perihelion per orbit is
\beq
\Delta \varpi_{\psi_{0}}=-\frac{3 \pi \mathcal M^4 (4+e^2)}{\psi_0  n^{2} a^6  (1-e^2)^3}; \label{eq:precpsi0}
\eeq
thus, the secular precession of $\varpi$, reintroducing physical units, is
\beq\left\langle\dot\varpi\right\rangle_{\psi_0}\doteq \Delta\varpi \left(\frac{n}{2\pi}\right)=-\frac{3 (GM)^4 (4+e^2)}{2\psi_0 c^6 n a^6  (1-e^2)^3}=-\frac{3 (G M)^{7/2}(4+e^2)}{2\psi_0 c^6  a^{9/2}  (1-e^2)^3}.\label{preces}\eeq
Note that eq. (\ref{preces}) is analytically different from the gravitoelectric $\mathcal{O}(c^{-2})$ Einstein precession which is proportional to $(GM)^{3/2}a^{-5/2}(1-e^2)^{-1}$.
\subsection{Alternative derivation of the perihelion precession} \label{ssec:alt}
Here we will  derive the perihelion precession due to HL gravity with an alternative approach based on the Lagrange planetary equations\cite{Ber} valid when the perturbation considered can be derived from a potential function.
Firstly, it is required to compute the average over one orbital revolution of the correction $\Delta U$ to the Newtonian  potential $U=-GM/r$ responsible of the perturbation.

In our case, from the expression of $h_{00}$ in eq. (\ref{eq:metric3}) it follows
\beq \Delta U_{\psi_0}= \frac{(GM)^4}{\psi_0 c^6 r^4}.\eeq
It is straightforward to obtain
\beq \mathcal{R}_{\psi_0}\doteq\left\langle \Delta U\right\rangle_{\psi_0}=\frac{(GM)^4(2+e^2)}{2\psi_0 c^6 a^4(1-e^2)^{5/2}}.\label{potz}\eeq

The Lagrange equations of the variation of $\omega$ and $\Omega$ due to a generic perturbation
are
\beq \left\langle\dot\omega\right\rangle=\frac{1}{n a^2 \sqrt{1-e^2}}\left[ \left(\frac{1-e^2}{e}\right)\frac{\partial \mathcal{R}}{\partial e}-\frac{\cos I}{\sin I}\frac{\partial \mathcal{R}}{\partial I}\right],\eeq
\beq \left\langle\dot\Omega\right\rangle = \frac{1}{n a^2 \sqrt{1-e^2}\sin I}\frac{\partial \mathcal{R}}{\partial I}.\eeq
Since $\mathcal{R}_{\psi_0}$ does not contain the inclination $I$, we simply have
\beq \left\langle\dot\varpi\right\rangle_{\psi_0}=\frac{\sqrt{1-e^2}}{n e a^2 }\frac{\partial \mathcal{R}_{\psi_0}}{\partial e},\label{ratez}\eeq
which yields just eq. (\ref{preces}).

\section{Confrontation with the observations}\label{paragone}
The expression of eq.(\ref{preces}) can be compared to the latest determinations of the corrections $\Delta\dot\varpi$ to the standard Newtonian/Einsteinian periehlion precessions of some planets of the solar system recently estimated by contrasting large planetary data sets of various types with accurate dynamical force models of (almost) all known Newtonian and general relativistic effects\cite{Pit1,Pit2,Pit10}. In view of the relatively short orbital periods of some of the planets considered with respect to the time interval spanned by the observational data set used, i.e. about one century, the estimated $\Delta\dot\varpi$ can be meaningfully compared to the predicted secular, i.e. averaged over one orbital revolution, precessions. Latest results obtained by Pitjeva\cite{Pit10} are shown in Table \ref{tavola_inner} and Table \ref{tavola_outer}; they are based on the EPM2008 ephemerides constructed from a data set of 550000 observations ranging from 1913 to 2008. It is interesting to note that she estimated corrections to the perihelion precessions of Neptune and Pluto as well, although the available observations do not yet cover  full orbital revolutions for them.

%
\begin{table}[ph]
\tbl{Estimated corrections $\Delta\dot\varpi$, in arcsec cty$^{-1}$, to the standard Newtonian/Einsteinian secular, i.e. averaged over one orbital revolution, precessions of the longitude of the perihelia of the inner planets obtained with the EPM2008 ephemerides\protect\cite{Pit10}. The quoted errors are not the mere formal, statistical ones. The corrections $\Delta\dot\varpi$ account for any un-modelled dynamical effects, classical (i.e. Newtonian/general relativistic) or not.}
{\begin{tabular}{@{}cccc@{}} \toprule
Mercury & Venus & Earth & Mars \\ \colrule
$-0.0040\pm 0.0050$ & $0.024\pm 0.033$ & $0.006\pm 0.007$ & $-0.007\pm 0.007$ \\
 \botrule
\end{tabular}\label{tavola_inner}}
\end{table}
\begin{table}[ph]
\tbl{Estimated corrections $\Delta\dot\varpi$, in arcsec cty$^{-1}$, to the standard Newtonian/Einsteinian secular precessions of the longitude of the perihelia of the outer planets obtained with the EPM2008 ephemerides\protect\cite{Pit10}. The quoted errors are not the mere formal, statistical ones. The corrections $\Delta\dot\varpi$ account for any un-modelled dynamical effects, classical (i.e. Newtonian/general relativistic) or not.\label{tavola_outer}
}
{\begin{tabular}{@{}ccccc@{}} \toprule
Jupiter & Saturn & Uranus & Neptune & Pluto\\ \colrule
$0.067\pm 0.093$ & $-0.010\pm 0.015$ & $-3.89\pm 3.90$ & $-4.44\pm 5.40$ & $2.84\pm 4.51$\\
\botrule
\end{tabular}}
\end{table}

Since the corrections to the standard perihelion precessions of Table \ref{tavola_inner}-Table \ref{tavola_outer} are statistically compatible with zero, attributing them to the action of the HL acceleration of eq. (\ref{acce}) implies that the maximum value of $\psi_0$ is infinity. Lower limits to it can be obtained from
\beq |\psi_0|\geq \left|-\frac{3 (G M)^{7/2}(4+e^2)}{2[\Delta\dot\varpi\pm\delta(\Delta\dot\varpi)]_{\rm max} c^6  a^{9/2}  (1-e^2)^3}\right|.\eeq
They are shown in Table \ref{tavola_psi_inner} for the inner planets and Table \ref{tavola_psi_outer} for the outer planets.
%
%
\begin{table}[ph]
\tbl{Lower bounds on $|\psi_0|$ obtained from the corrections $\Delta\dot\varpi$ to the standard perihelion precessions of  the inner planets of Table \ref{tavola_inner} by using the predicted rate of eq. (\ref{preces}).\label{tavola_psi_inner}
}
{\begin{tabular}{@{}cccc@{}} \toprule
Mercury & Venus & Earth & Mars \\ \colrule
$8\times 10^{-12}$ & $6\times 10^{-14}$ & $6\times 10^{-14}$ & $8\times 10^{-15}$ \\
\botrule
\end{tabular}}
\end{table}
\begin{table}[ph]
\tbl{Lower bounds on $|\psi_0|$ obtained from the corrections $\Delta\dot\varpi$ to the standard perihelion precessions of  the outer planets of Table \ref{tavola_outer} by using the predicted rate of eq. (\ref{preces}).\label{tavola_psi_outer}
}
{\begin{tabular}{@{}ccccc@{}} \toprule
Jupiter & Saturn & Uranus & Neptune & Pluto \\ \colrule
$3\times 10^{-18}$ & $1\times 10^{-18}$ & $2\times 10^{-22}$ & $2\times 10^{-23}$ & $8\times 10^{-24}$\\
 \botrule

\end{tabular}}
\end{table}
The larger value is $8\times 10^{-12}$ and comes from Mercury. As an a-posteriori check of the consistency of our perturbative results with the assumption of eq. (\ref{condiz}), we note that our lower bounds on $\psi_0$ are always larger than $4({\mathcal{M}}/{r})^3$ for all the planets considered.

Let us note that $\psi_0=8\times 10^{-12}$ (see Ref.~\refcite{Tib}) yields for Mercury a precession of $-0.07007$ arcsec cty$^{-1}$, which is quite different from the standard Einstein precession of $+42.98$ arcsec cty$^{-1}$.

Finally, we mention  that since $\vec{A}_{\psi_0}$ is a radial acceleration, although directed outward the Sun and not spatially uniform, it may be interesting to see if our limiting values for $\psi_0$ may, at least, explain the magnitude of the anomalous Pioneer acceleration ($A_{\rm Pio}=8.74\times 10^{-10}$ m s$^{-1}$) experienced by the two twin probes after 20 AU\cite{Pio}.
The answer is negative. Indeed, the upper bounds on $\vec{A}_{\psi_0}$, corresponding to the lower bounds on $\psi_0$ obtained from Uranus, Neptune and Pluto orbiting in the spatial regions in which the Pioneer anomaly manifested itself in its presently known form, are smaller by about one order of magnitude. They are shown in Table \ref{tavola_Pio}.
%
\begin{table}[ph]
\tbl{Upper bounds of the HL acceleration of eq. (\ref{acce}) at heliocentric distances of Uranus, Neptune and Pluto computed with the lower bounds on $\psi_{0}$ according to Table \ref{tavola_psi_outer}.\label{tavola_Pio}
}
{\begin{tabular}{@{}ccc@{}} \toprule
20 AU & 30 AU & 40 AU \\
 \colrule
$5.4\times 10^{-11}$ m s$^{-2}$ & $5.5\times 10^{-11}$ m s$^{-2}$ & $3\times 10^{-11}$ m s$^{-2}$\\
\botrule
\end{tabular}}
\end{table}
\section{Discussion and conclusions}\label{concludi}
In this paper we considered the spherically symmetric  KS solution, which is a solution of HL gravity, obtained relaxing both the detailed balance and projectability conditions. The Newtonian and lowest order post-Newtonian limits of KS solution coincide with those of GR, and we focused on the first corrections to the GR behaviour, in order to evaluate the impact of these corrections on the solar system orbital motions and use the available data to constrain this particular solution of HL gravity. Actually, our results apply as well to arbitrary perturbations of the Schwarzschild proportional to $1/r^{4}$.

After having analytically calculated the  pericentre precession of a test particle averaged over one orbital revolution with both the Gauss and Lagrange perturbative schemes, we compared our theoretical prediction to the corrections to the standard Newtonian/Einsteinian perihelion precessions of all the planets of the solar system, recently estimated by E.V. Pitjeva by fitting  the dynamical models of the EPM2008  ephemerides to 550000 observations of several types spanning the time interval 1913-2008. The upper bounds on the estimated corrections to the perihelion precessions of all the planets, taken singularly one at a time, yield lower bounds on $\psi_0$. The largest one occurs for Mercury, being of the order of  $10^{-12}$, while Pluto yields the smallest one which is about $10^{-24}$. Concerning Mercury, the corresponding lower bound for $\psi_0$ yields a retrograde perihelion precession of at most $-0.07007$ arcsec cty$^{-1}$, quite different from the standard general relativistic one. The lower bounds on $\psi_0$ from Uranus, Neptune and Pluto, moving where the Pioneer anomaly manifested itself in its presently known form, yield upper bounds on the corresponding HL acceleration one order of magnitude smaller than the anomalous Pioneer acceleration of $8.74\times 10^{-10}$ m s$^{-2}$.

If and when other teams of astronomers will estimate their own corrections to the standard rates of the planetary of perihelia, it will be possible to repeat the present analysis. A further, complementary approach that could be followed, although very time-consuming, consists in re-processing all the planetary data sets with the fully modeled  KS metric and looking at the values of the estimated parameters, including also $\psi_0$ itself.

\section*{Acknowledgements}
We thanks some anonymous referees for their useful and constructive remarks.

\end{document}